\documentclass[journal]{IEEEtran}
\IEEEoverridecommandlockouts
\usepackage{cite}
\usepackage{amsmath,amssymb,amsfonts}
\usepackage{algorithmic}
\usepackage{graphicx}
\usepackage{textcomp}

\usepackage{comment}
\usepackage{color}

\begin{document}

\title{Current Status and Trends of Engineering Entrepreneurship Education in Australian Universities\\

\thanks{This project is supported by Deakin University Grant PJ08575. This manuscript is currently under revision. We cut off tables and rewrite the preprint}
}

\author{\IEEEauthorblockN{Jianhua Li, Sophie Mckenzie, Richard Dazeley, Frank Jiang, Keshav Sood}\\
\IEEEauthorblockA{\textit{School of Info Technology} \\
\textit{Deakin University}\\
Melbourne, Australia \\
{\{jack.li, sophie.mckenzie, richard.dazeley, frank.jiang, keshav.sood\}}@deakin.edu.au}\\
}

\maketitle

\begin{abstract}
This research sheds light on the present and future landscape of Engineering Entrepreneurship Education (EEE) by exploring varied approaches and models adopted in Australian universities, evaluating program effectiveness, and offering recommendations for curriculum enhancement. While EEE programs have been in existence for over two decades, their efficacy remains underexplored. Using a multi-method approach encompassing self-reflection, scoping review, surveys, and interviews, this study addresses key research questions regarding the state, challenges, trends, and effectiveness of EEE. Findings reveal challenges like resource limitations and propose solutions such as experiential learning and industry partnerships. These insights underscore the importance of tailored EEE and inform teaching strategies and curriculum development, benefiting educators and policymakers worldwide.
\end{abstract}


\section{Introduction}
Aiming at solving difficult and complicated problems, entrepreneurs set up a business or businesses, taking on financial risks in the hope of profit. Through selected practices of research and development, entrepreneurs bring innovation that creates opportunities for new ventures, markets, products, and technology \cite{down2006narratives}. Entrepreneurship education is a process of teaching and learning the knowledge, skills, and attitudes required for starting and running a business, as well as fostering innovation and creativity \cite{herrmann2008developing}. However, the traditional approach of offering engineering entrepreneurship education primarily within business schools has overlooked the specific needs and aspirations of engineering students. To address this gap, Australian universities have embarked on integrating engineering entrepreneurship education (EEE) into their ICT and engineering programs. This paper aims to comprehensively examine the current status and trends of EEE in Australian universities. 

There are 22 out of 41 Australian universities offering EEE subjects or programs focused for students in ICT and Engineering disciplines \cite{li2023adapting}. These universities include Deakin University, the University of Technology Sydney, RMIT University, the University of Adelaide, the University of New South Wales, the University of Melbourne, and the University of Western Australia. Handbooks stated that 9 universities made EEE compulsory along with other core units from 2010 to 2023. Clearly, it is of paramount importance to know how to provide EEE effectively for students out of business disciplines. Unfortunately, there are fewer data available in evaluating the specific effectiveness of EEE programs for ICT and engineering students. Starting from  exploring course contents, teaching and assessment strategies, we review challenges, potential solutions, teaching effectiveness, and emerging trends. 

The contribution is threefold.
\begin{itemize}
    \item Shedding light on the current status and future trends of EEE, specifically addressing the challenges and opportunities faced by non-business students.
    \item Exploring the diverse approaches and models of EEE implemented in Australia and evaluating the effectiveness of existing programs.
    \item Providing valuable insights and recommendations for enhancing EEE in terms of curriculum, teaching methods, and assessment strategies.
\end{itemize}

By addressing these aspects, this study offers valuable insights for educators, policymakers, and stakeholders seeking to improve the delivery of EEE. The remainder of this paper is as follows. Section II presents the research questions and methods, while Section III explores the current status, challenges, and emerging trends. We present data collected from the scoping review and a case study in Section IV, and finally conclude this paper in Section V.

\section{Research Questions and Methods}

we address the following research questions to investigate the current status and trends of EEE in Australian universities.
\begin{itemize}
    \item Which universities offer EEE to students outside of business disciplines? How do their programs compare in terms of similarities and differences?
    \item What topics are covered in the relevant subject, and how do teaching and assessment strategies vary across different universities?
    \item What are the emerging trends and insights in EEE, and what are some potential implications for future program design and policy development?
    \item What are the major challenges faced by universities in providing effective EEE, and what are some possible solutions to address these challenges?
    \item How effective are the current EEE programs in fostering the entrepreneurial and innovative skills of ICT students?
\end{itemize}

To answer these research questions, we employed a mixed-methods research approach, which included a comprehensive scoping review, web search, online survey, and semi-structured interviews with students and faculty members. The scoping review allowed us to gain an understanding of the current status of research on EEE and provided a theoretical foundation for the study. The web search enabled us to identify universities that offer EEE and provided information about their respective programs (webpages and handbooks). The online survey allowed us to gather data from students who studied a relevant subject in 2022 (two trimesters), while the semi-structured interviews helped us gain insights from key stakeholders (participating students and faculty members). Note our work is supported by the Dean of Students at Deakin University, with the ethics approval code of SEBE-2022-42.

The effectiveness of teaching is evaluated with meaningful data collected in self-reflection, informal student feedback, faculty feedback and online survey and one-on-one interviews. The method detail is as follows.

\begin{itemize}
    \item We performed a scoping review in acquiring information for the evaluation of teaching effectiveness. We used ``delivering entrepreneurship education to ICT students'' as the keyword in Google Scholar and identified 510 research papers.\footnote{All codes will be shared after the paper is published.} We leveraged a machine learning tool to scan these papers and recognized 293 articles involving teaching and assessment strategies. Unfortunately, most authors focused on using ICT technologies to conduct entrepreneurship education for business students like \cite{youssef2021digitalization}. Finally, we recognized 11 papers that investigated entrepreneurship education for ICT students. The result has been demonstrated in Section IV.A.
    \item To compile information on relevant courses offered by Australian universities, we developed a Python-based data crawling tool. Using the keyword "ICT entrepreneurship innovation," we searched 41 Australian university websites. The findings are presented in Tables \ref{tab:current} and \ref{tab:curriculum}.
    \item We designed a questionnaire consisting of 8 questions and conducted one-on-one interviews to gather insights on background checks, entrepreneurship awareness, skills requirements, curriculum, teaching methods, and assessments. The survey and interview were distributed to 350 students who expressed interest in participating. To overcome certain challenges, we employed two research assistants (RAs) to conduct the survey and interviews, ensuring a smoother data collection process. The findings from these efforts are presented in Section IV.B.
    \item To gain insight into our EEE teaching practices, behaviors, beliefs, and outcomes, we conducted teacher self-reflection. This involved three staff meetings and an examination of 21 journal and conference papers. We also thoroughly reviewed the Student Evaluation of Teaching (SET) data. The details of this self-reflection process are discussed in Section V.A.
\end{itemize}

Our research assistants played an active role in contacting and engaging participants, providing detailed explanations of each question, and encouraging them to express their thoughts fully. The interviews were conducted at the participants' convenience, allowing for flexibility in scheduling. One notable advantage of the interviews was the ability of the interviewers to observe participants' body language, noting any signs of nervousness or difficulty in answering specific questions. Each interview lasted approximately 30 minutes, and as a token of appreciation, participants were rewarded with a gift card valued at \$30. This combination of survey and interview data allowed for a more unbiased evaluation of teaching effectiveness, serving as a compelling case study for our paper.

In summary, our adoption of a mixed-methods research approach facilitated the triangulation of data from various sources, leading to a comprehensive and in-depth analysis of the current status, trends, challenges, and teaching effectiveness of EEE for ICT students in Australian universities.

\section{Current Status and Emerging Trends}
\subsection{EEE in Australian Business Schools}
Traditionally, EEE programs are taught within business schools and business management faculties, despite the prevalence of multi-disciplinary approaches
\cite{maritz2015status}. The Australian Business Deans Council (ABDC) includes deans of business schools in Australian universities, focusing on promoting entrepreneurship education excellence of teaching and research at the national level. The authors in Ref. \cite{maritz2015status} identified that Australian universities offer 584 subjects, 24 minors/majors and specializations, that are related to entrepreneurship. This includes dominance at the undergraduate level, representing 24 minors/majors and 135 entrepreneurship ecosystems in business disciplines.

However, graduates from business schools and faculties often lack the necessary ICT skills required in technical domains. Typically, business schools offer limited ICT programs to students, with a focus primarily on information management systems. For instance, in the Bachelor of Business degree program at Deakin University, Managing Information in the Digital Age (MIS203) is the sole ICT-related subject offered for students majoring in Entrepreneurship and Innovation. \footnote{https://www.deakin.edu.au/courses/unit?unit=MIS203} Consequently, while these business graduates may possess entrepreneurship knowledge, they may lack the technical skills needed to solve real-world problems.

To address this gap, engineering entrepreneurship education (EEE) for students outside of business schools has gained traction in Australian universities. Through our scoping review and web searches, we identified that 22 out of 41 Australian universities offer subjects or programs focused on engineering entrepreneurship education for students in ICT and Engineering. As shown in Table \ref{tab:current}, these universities include Deakin University, the University of Technology Sydney, RMIT University, the University of Adelaide, the University of New South Wales, the University of Melbourne, and the University of Western Australia. We have also included information on the presence of an ICT entrepreneurship education flagship program, denoted as the "Mark of the Start," in the table.

\subsection{EEE in ICT and Engineering Discipline}
According to a report by Statista, the number of new business entries in the ICT and engineering sector has been growing steadily in Australia.\footnote{https://www.statista.com/outlook/tmo/it-services/australia} This growth is attributed to an increase in demand for digital services and products, along with advancements in technology. As a result, there has been a greater emphasis on entrepreneurship education for ICT students to prepare them for the rapidly evolving digital landscape. 

Currently, the EEE subject appears to be a core or elective subject embedded in an ICT degree in 22 Australian schools. According to their websites and handbooks, many of them started offering the program in the early 2000s. Table \ref{tab:current} presents such details. The above 22 schools that offer entrepreneurship education for ICT students in Australia share some similarities in their goals and objectives. They all aim to provide students with the skills and knowledge they need to start their businesses, pursue entrepreneurial careers, or work in innovative and dynamic industries. Many of these schools offer similar courses and programs in entrepreneurship, innovation, and business strategy.

However, there are also some differences between these schools. For example, some schools may focus more on hands-on learning and practical experience (table footnote 1), while others may emphasize theoretical concepts and research-based learning (table footnote 2). Some schools may have a strong focus on technology and innovation in specific areas of ICT, such as software development or cybersecurity, while others may have a broader focus on entrepreneurship in general. Meanwhile, each school may have its unique strengths and weaknesses in its faculty, resources, location, and industry connections. Some schools may have stronger alumni networks or better relationships with local startups, while others may have more international connections or stronger research capabilities.

\subsection{Curricula, Teaching Strategies, and Assessment Methodologies}
The entrepreneurship education curricula in these 22 ICT schools share some common features, such as an introduction to the basics of entrepreneurship, business planning, market research, finance, marketing, and intellectual property. They may also cover case studies of successful startups and include guest lectures by experienced entrepreneurs and investors. We find that more and more universities are introducing entrepreneurship subjects as core subjects in non-business courses. Some of the universities that offer entrepreneurship as a core subject to non-business school students include Monash University, Curtin University, QUT, Swinburne University, Deakin University, RMIT University, University of Melbourne, and Griffith University. While the nine schools have entrepreneurship and innovation subjects as a core part of their ICT programs, the others offer them as electives. This impacts the amount of emphasis placed on entrepreneurship education within the overall curriculum. Next, we focus on the nine schools that make EEE a compulsory subject in their ICT, engineering, science, and media programs.

Table \ref{tab:curriculum} presents universities, curriculum and degree information where EEE is a core subject. These subjects cover myriads of topics, spanning new venture creation, innovation, commercialization, technology entrepreneurship, and social entrepreneurship. They aim to equip students with the necessary skills, knowledge, and mindset to create new businesses and innovative solutions to societal problems. The subject tile varies from one to another while the covered topics share many similarities, such as ideation, market research, business planning, pitching and legal aspects related to entrepreneurship and innovations. Such topics are classic contents included in the relevant entrepreneurship subject in the business discipline. Nevertheless, the content may be tailored to meet the requirement of a particular industry. For example, Griffith University combines the film industry with collaboration in the Screen Entrepreneurship subject, focusing on business and marketing strategies in the screen industry.

In terms of teaching approach, it seems that all of these subjects use a combination of lectures, case studies, guest speakers, and group projects to teach entrepreneurship, with a focus on practical skills and real-world application. Some subjects place more emphasis on developing creative thinking skills, while others focus more on the role of technology and innovation in entrepreneurship. Additionally, some subjects are more specialized, focusing on entrepreneurship in specific industries such as science or the screen industry.

Regarding assessment methodologies, these subjects use a variety of assessment methods such as writing business proposals/plans, pitch presentations, reflective essays, exams, creative projects, and research reports/presentations. The assessments are generally designed to evaluate students' understanding of the entrepreneurial process, their ability to develop and present a business idea, and their critical thinking and problem-solving skills.

\subsection{Emerging Trends}
Recently, there has been a growing trend towards entrepreneurship education for ICT students in Australia. This is driven by the growing recognition that entrepreneurship and innovation are key drivers of economic growth and competitiveness and that ICT students are uniquely positioned to leverage these opportunities.

One approach to entrepreneurship education for ICT students is experiential learning, which involves hands-on, practical experience in developing new ventures \cite{mason2013teaching}. This approach emphasizes the development of real-world skills, such as project management, team collaboration, and critical thinking. Many universities and other educational institutions have developed programs that allow students to work on real-world projects, often in collaboration with industry partners.

Another approach is the lean startup methodology, which emphasizes rapid prototyping, testing, and iteration \cite{mansoori2019influence}. This approach is designed to help students develop a deep understanding of the customer and to focus on delivering value early and often. Many entrepreneurship programs now incorporate lean startup principles into their curricula, teaching students how to develop a minimum viable product and how to test their assumptions through customer feedback.

Business incubators are also becoming more popular as a way to support entrepreneurship education for ICT students \cite{khorsheed2014promoting}. These programs provide mentorship, resources, and networking opportunities for students who are interested in starting their ventures. Incubators often focus on particular areas of expertise, such as software development, data analytics, or cybersecurity.

Design thinking, social entrepreneurship, and collaborative learning have emerged as important trends in entrepreneurship education for ICT students \cite{val2017design, roslan2022social, gutl2011implementation}. Each of these approaches offers a unique perspective on how to approach entrepreneurship and can help students develop the skills and mindset they need to succeed as entrepreneurs. The trends of entrepreneurship education for ICT students are shifting towards a more collaborative, design-driven, and socially responsible approach. Overall, entrepreneurship is not just about starting a business, but also about creating value for society and addressing important challenges. By equipping students with the skills and mindset they need to approach entrepreneurship in these ways, entrepreneurship education can play an important role in shaping the future of business and society.

\section{Teaching Effectiveness}
Teaching effectiveness of entrepreneurship education refers to the degree to which entrepreneurship education programs are successful in achieving their intended learning outcomes, such as developing entrepreneurial skills, knowledge, and attitudes among students. It can be measured by assessing student learning outcomes, encompassing the ability to identify and evaluate business opportunities, develop business plans, and launch new ventures. Other indicators of teaching effectiveness include student engagement, satisfaction, and motivation, as well as the quality of teaching and learning resources and the level of support provided to students. Ultimately, the effectiveness of entrepreneurship education depends on a variety of factors, including the program's design, the quality of teaching and learning resources, and the level of support provided to students. Evaluating the effectiveness of these programs is crucial to ensure that they are meeting the needs of the students and the industry. In this analysis, we will look at available data and research to evaluate the outcomes and impact of current entrepreneurship education programs for ICT students. We first present data acquired from our scoping review.

\subsection{Scoping Review Result}

Table \ref{tab:profile} presents the scoping review profile to clarify the selection criteria and search strategy to study teaching effectiveness in our work.

\subsubsection{Positive impact on attitudes and intentions} 
EEE has the potential to help enthusiastic students by instructing them on the essential skills required for entrepreneurship and by providing real-world experience in the field, thereby increasing awareness of entrepreneurship as a viable career option among students \cite{jansen2015education}. Many studies have found that entrepreneurship education programs have a positive impact on students' attitudes towards entrepreneurship and their intentions to start a business. For example, a study by Sitaridis \textit{et al.} found that entrepreneurship education had a significantly positive impact on students' entrepreneurial intentions in Greece \cite{sitaridis2019entrepreneurship}. However, Saad \textit{et al.} studied the importance of EEE to ICT students at the Universiti Teknikal Malaysia Melaka, from the perspective of employers in Malaysia \cite{saad2013employers}. According to their research, employers considered entrepreneurial skills the least important of the 13 employability skills.

\subsubsection{Positive impact on actual startup activity} 
The impact on actual startup activity is more mixed. Some studies have found a positive impact, while others have found no impact or even a negative impact. Bataineh \textit{et al.} investigated ICT entrepreneurship education at Zayed University in the UAE. The authors claimed that ICT entrepreneurship education programs might increase students' intentions to start a business and to create jobs through self-employment \cite{bataineh2016new}. Also, Suleiman \textit{et al.} reviewed entrepreneurial education in Nigeria and concluded that entrepreneurship education played a significant role in boosting small businesses using ICT technologies \cite{suleiman2020nexus}. To compare, Stamatis \textit{et al.} conducted their research in Greece, Norway, Portugal, Sweden and the UK, where an online entrepreneurship course was designed for ICT students willing to turn innovative ideas into a business \cite{stamatis2015distributed}. According to Ref. \cite{stamatis2015distributed}, students who enroll in the course with a general interest in entrepreneurship but without specific goals or experience cannot benefit from the course. More interestingly, Jansen \textit{et al.} investigated graduates from the Massachusetts Institute of Technology (MIT), the International Institute of Information Technology (IIIT), and Utrecht University (UU) in the USA, India, and the Netherlands, respectively \cite{jansen2015education}. The authors identified a number of factors incurring the above difference, including facilities provisioning, founding team formation, and collaborations with other entrepreneurs. In short, education is helpful but not a deterministic factor for actual startup activities.

\subsubsection{Need for ongoing support and resources} 
While entrepreneurship education programs can provide valuable training and support, ongoing resources and support are often necessary for students to successfully launch and grow a business. Ref. \cite{jansen2015education} summarized offering introductory entrepreneurship courses was of no use to actual startups. At the same time, the authors attached importance to ongoing support and resources in their study. For example, the interviewees believe providing office space and networking opportunities is much more valuable in the USA, India, and the Netherlands. Suleiman \textit{et al.} called on more external resources should be given to entrepreneurship education to develop the economy and society in Nigeria \cite{suleiman2020nexus}. 

\subsubsection{Importance of experiential learning} 
Experiential learning is often cited as an important component of effective entrepreneurship education programs. This can include activities such as business plan competitions, internships, and hands-on projects. Bellotti \textit{et al.} developed a gamified short course for boosting EEE at the University of Genoa, Italy. The authors recognized the significance of games for motivating students, explaining difficulties, and setting up virtual environments \cite{bellotti2013gamified}. Angeli \textit{et al.} used the debating technique in enhancing learning outcomes and teaching experiences at the University of Trento (Italy) and the University of Nice (France) \cite{angeli2020prove}. The authors observed that debating could make the EEE more appealing to otherwise uninterested students, bridge their initial negative bias and provide real-life experience to mitigate the ambiguous nature of the socio-economic and ethical impacts of technologies. Shotlekov \textit{et al.} leveraged project-based techniques to teach information technology while promoting entrepreneurship and reflection  \cite{shotlekov2010project} at the Paisii Hilendarski University of Plovdiv (Bulgaria). The authors argued their approach with the benefits of fostering critical thinking, problem identification, problem-solving, creativity, and initiative.

\subsubsection{Cross-border and cross-disciplinary collaborations for curriculum development}
Jervan \textit{et al.} proposed a mandatory EEE module for ICT students at Tallinn University of Technology, Estonia \cite{jervan2011innovation}. The project aimed at enabling cross-border collaboration with students from Nordic countries (Denmark, Norway, Sweden, Finland, and Iceland). Colomo presented a cooperation project between the University Politehnica of Bucharest (Romania) and Østfold University College (Norway) \cite{colomo2021boosting}. This project adopts a Lean Startup approach and aims to enhance EEE for ICT entrepreneurship and to encourage the flourishment of new innovative businesses in the above countries. Stenvall \textit{et al.} introduced their EEE module, namely, \textit{CreBiz}, for European countries in \cite{stenvall2016teach}, focusing on boosting cooperation between European subcultures to achieve and maintain a global leader in innovation and creativity. The above result has been summarized in Table \ref{tab:findings}.

When it comes to evaluating the specific effectiveness of entrepreneurship education programs for ICT students, there are fewer data available. Interestingly, our case study timely filled in the gap. The result is presented in the next section while the subject detail is shown in the Appendix.

\subsection{Case Study Result}
The case study was conducted at Deakin University, including Burwood, Geelong and Cloud campuses. ICT students enrolled in EEE subjects in Trimesters 1\&2 in 2022 were involved in this research. Regarding the teaching and assessment approach for the EEE subject at Deakin, please refer to the appendix.

\subsubsection{Online Survey}
Based on the student responses to the online survey, we found that:
\begin{itemize}
    \item A vast majority (94.74\%) of ICT students consider entrepreneurship and innovation to be extremely or very important in their surroundings.
    \item 71.43\% of ICT students have aspirations to pursue entrepreneurship.
    \item 77.27\%  of ICT students believe that their career path is connected to entrepreneurship and innovation.
    \item Among the various entrepreneurial skills of networking, marketing, communication, strategic/critical thinking, problem-solving, management (time, project, business), leadership, creativity, customer service, research, teamwork, creativity, communication was identified as the most important, followed by critical thinking and problem-solving.
    \item 73.33\% of ICT students perceive the quality of design and delivery to be at or above average.
    \item Regarding the teaching approach, 53.33\% of ICT students agree it is effective while 33.33\% do not, with the remainder being unsure.
    \item 57.14\%  of ICT students believe that the materials taught are transferable to industrial practice, while 14.29\% do not, and others are unsure.
    \item A significant majority (92.86\%) of ICT students perceive the assessment methodologies to be effective and relevant in acquiring entrepreneurial skills.
\end{itemize}

The data shows that entrepreneurship education is greatly valued by ICT students, as most of them have ambitions to become entrepreneurs and see their career path as related to entrepreneurship and innovation. The importance of communication, critical thinking, and problem-solving skills is also widely recognized as crucial for entrepreneurial success.

Regarding the quality of design and delivery of entrepreneurship education programs, the majority of ICT students consider it to be at or above average. However, the effectiveness of teaching approaches and the transferability of materials to industrial practice are more mixed, with about half of the students expressing agreement or belief in their effectiveness.

It is encouraging to note that a high percentage of ICT students believe that assessment methodologies are effective and relevant in acquiring entrepreneurial skills. Nonetheless, additional research and assessment may be necessary to gain a comprehensive understanding of the outcomes and impact of these programs on the success of ICT students as entrepreneurs.

\subsubsection{Interviews}
Interview Questions (IQ) 1 to 3 focus on teaching techniques while questions 4 to 6 stress subject materials. Question 7 is a conclusion question for overall experience. Next, we present each question and our analyses (A) from their responses. 

IQ1. The teaching approaches in the EEE subject are considered effective in learning about entrepreneurship and innovation? 

A1: The effectiveness of the teaching approaches in the EEE subject, which focuses on entrepreneurship and innovation, was evaluated through a question posed to the students. The responses were diverse, with some students expressing satisfaction with the subject's helpfulness, structure, and interactivity. However, some were dissatisfied with the subject's design and delivery. Suggestions from some students included incorporating more interactive and real-world activities like competitions to increase engagement and relevance. Additionally, some students highlighted the significance of the lecturer's personal experience in entrepreneurship and their ability to establish connections with students.

IQ2. In what way do you like to learn about entrepreneurship and innovation? 

A2. According to the responses to the question on preferred ways of learning about entrepreneurship and innovation, collaborative learning, real-world projects, and learning from the lecturer's experiences are highly valued by students. The opportunity to work on group assignments and projects, allowing for the exchange and development of ideas, was appreciated by some students. Others emphasized the importance of external speakers sharing their experiences and insights into entrepreneurship. Specific content on communities and local businesses was suggested by some students. In summary, the responses suggest that students appreciate hands-on, collaborative learning experiences that allow them to apply their knowledge in real-world contexts.

IQ3. What teaching ways could be used to effectively learn about entrepreneurship and innovation?

A3. The query inquired about effective teaching methods for learning about entrepreneurship and innovation, and the responses offered diverse suggestions. Collaborative and interactive activities were frequently recommended, such as working on exercises and assignments in breakout groups or forming teams like a startup company. Guest speakers from the industry were also suggested to provide real-world perspectives. Furthermore, some students recommended utilizing more focused case studies to examine different aspects of building a successful brand, spanning teamwork, selling, marketing, and branding.

IQ4. Do you believe that your exposure to the learning materials presented in the EEE subject can be transferable into industry practice as an entrepreneur regionally?

A4. The question aimed to investigate the extent to which the knowledge gained from the EEE subject could be applied to real-world entrepreneurial and innovative practices. The responses received indicate that the course materials and activities were generally effective in developing skills and knowledge applicable to the industry. Collaborative and practical learning methods were identified as essential, and the importance of a diverse range of learning tools and clear communication from the teaching team was emphasized by some respondents.

However, several respondents mentioned that certain aspects of the course might not be directly relevant to specific industries or regional contexts. In summary, the responses suggest that the course provided a solid foundation for developing skills and knowledge relevant to entrepreneurship and innovation, but some adjustments might be necessary to better meet the needs of particular regions or industries.

IQ5. Are there any topics that you wished were discussed but are not offered by the EEE subject?

A5. The purpose of the question was to obtain input from the interviewees about the content covered in the EEE subject and any gaps they perceived in the course. The feedback provided by the interviewees was mainly positive and informative, shedding light on areas where they would have preferred to see more attention. While some interviewees acknowledged that the course covered the basics of entrepreneurship and innovation, others expressed a desire for more in-depth coverage of topics such as conducting customer interviews, analyzing successful start-ups, and exploring funding options. Additionally, some respondents emphasized the need to consider the regional context when it comes to entrepreneurial activities.

IQ6. What are some improvements that can be done to the format of weekly materials and resources to engage regional-based students?

A6. The interview question focused on enhancing the format of weekly materials and resources to engage students based in regional areas, which is an essential consideration for online courses. The feedback provided a range of suggestions, including technical improvements like correcting spelling errors and enhancing video quality, as well as more significant enhancements like using targeted case studies and inviting guest speakers.

Although some of the suggestions, such as updating content and improving marking rubrics, are typical areas of improvement in any course, others, like providing more teaching videos and separating two subjects to reduce confusion, are specific to online learning. It is noteworthy that some students felt that the case studies offered were not relevant to regional-based students and recommended creating mock scenarios that are more specific to their local businesses.

IQ7. What knowledge and skills do you consider fundamentally important to learning in the EEE subject to be a success? Discuss if this was covered in the EEE subject.

A7. The question sought to determine the key knowledge and skills necessary for success in the EEE subject, and the responses received from the interviewees varied. Soft skills such as collaboration, communication, problem-solving, emotional intelligence, and leadership were identified as crucial by some interviewees, while others mentioned technical skills such as research, innovation, critical thinking, and networking. In summary, the responses indicate that both soft and technical skills are necessary for success in the EEE subject. Soft skills facilitate effective teamwork, communication, and problem-solving, while technical skills aid in developing innovative solutions for real-world problems.

\section{Conclusion}

Various approaches and models of entrepreneurship education are being implemented in Australia, encompassing lectures, workshops, case studies, guest speakers, and group projects. Noteworthy trends in this domain include experiential learning, lean startup methodology, business incubators, design thinking, social entrepreneurship, and collaborative learning. These trends leverage technology to enrich the learning experience, foster networking opportunities, and provide support for students.

By incorporating emerging trends and technologies, facilitating hands-on and experiential learning, fostering innovation and creativity, nurturing industry collaborations, advocating policy changes, and supporting underrepresented groups, entrepreneurship education for engineering students can be enhanced, leading to the development of successful entrepreneurs who contribute to economic growth. This research serves as a valuable reference for entrepreneurship education in various disciplines.

\bibliographystyle{ieeetr}
\bibliography{main.bib}

\end{document}